\documentclass[10pt,letterpaper]{article}
\usepackage{opex3}

\begin{document}

\title{On-chip spectro-detection for fully integrated coherent beam combiners}

\author{Pierre Kern (a), Etienne Le Co\"arer (a), Pierre Benech (b)}

\address{a) Laboratoire d'Astrophysique de Grenoble, Universit\'e Joseph Fourier, CNRS, Grenoble, France; \\
b) Institut de Micro\'electronique d'Electromagn\'etisme et de Photonique, INPG-UJF-CNRS, Grenoble, France }
\email{Pierre.Kern@obs.ujf-grenoble.fr} 



\begin{abstract}
This paper presents how photonics associated with new arising detection technologies is able to provide fully integrated instrument for coherent beam combination applied to astrophysical interferometry. The feasibility and operation of on-chip coherent beam combiners has been already demonstrated using various interferometric combination schemes. More recently we proposed a new detection principle aimed at directly sampling and extracting the spectral information of an input signal together with its flux level measurement. The so-called SWIFTS demonstrated concept that stands for Stationary-Wave Integrated Fourier Transform Spectrometer, provides full spectral and spatial information recorded simultaneously thanks to a motionless detecting device.

Due to some newly available detection principles considered for the implementation of the SWIFTS concept, some technologies can even provide photo-counting operation that brought a significant extension of the interferometry domain of investigation in astrophysics . The proposed concept is applicable to most of the interferometric instrumental modes including fringe tracking, fast and sensitive detection, Fourier spectral reconstruction and also to manage a large number of incoming beams. The paper presents three practical implementations, two dealing with pair-wise integrated optics beam combinations and the third one with an all-in-one 8 beam combination. In all cases the principles turned into a pair wise baseline coding after proper data processing.
\end{abstract}




\section{Introduction}
\label{sec:intro}  
The implementation of the instruments dedicated to the coherent beam combination for stellar interferometry involves very complex optical setups.  Existing interferometric facilities that are available for astronomical observations deliver coherent beams coming from up to 8 telescopes. A limiting factor for the combination of all available beams of a given interferometer is this complexity and the induced high stability requirement. Using fiber optics and integrated optics devices one can address part of this concern \cite{Kern96} and efficient implementations on several instruments already provide noteworthy astrophysical results \cite{kraus05,monnier04, amb07}. Major interferometric facilities, like the European Very Large Telescope Interferometer (VLTI - Cerro Paranal - Chile) (4x8~m telescopes and 4x1.8m telescopes) or Chara in Mt Wilson-California (6x1m telescopes) are currently operating instruments that make usage of photonics components and consider for their next generation instruments more complex photonics devices. 

While we demonstrated in the past the efficiency of photonics devices for multi beam combination in single mode interferometry \cite{berger01}, we show here how its association with the concept of Stationary-Wave Integrated Fourier Transform Spectrometer (SWIFTS hereafter) \cite{LeCoarer07} allows us to consider a fully integrated instrument, including coherent beam combination, spectrometry and detection implemented on a single photonics device.
The association of SWIFTS evanescent detection principle and photonics beam combination brings a great potential for a fully integrated beam combiner instrument.

Section \ref{sec:techn} reminds the reader that the state of the art of the considered technologies as well as the main requirements for an interferometric beam combining instrument. According to the purposes of the considered instrument some specific requirements must be emphasized: high precision measurement of the fringe contrast, accurate phase measurements, spectral resolution capabilities, high dynamic capability on the photometric signal for the detection of faint object in the vicinity of bright stars, high imagery reconstruction capability using the largest number of telescopes, highest sensitivity to achieve observations on the fainter objects.
In this section we remind the reader also of the principle of the SWIFTS concept. Section \ref{sec:swifts} describes several possible implementations of fully integrated instruments. We address for each proposed concept its adequacy the purposes of some specific instruments. We show that using existing built-in blocks, the requirements for the interferometric instrumentation can be met using a single integrated instrument.

\section{Background and available technologies}
\label{sec:techn}

\subsection{Instruments for interferometry}
The requirements of an interferometric instrument must be adapted according to its main purposes. An interferometric fringe pattern has two meaningful observables, its contrast, or depth of modulation and its phase or location of the central fringe.

 \textbf{High precision contrast} measurement is required to perform an accurate reconstruction from the measured fringes, using the appropriate Fourier transform. Many instrumental biases strongly affect the fringe patterns and drastically increase the error on the reconstructed signal.
 
  \textbf{An accurate phase determination} is required for both astrometric applications, and to ensure an efficient detection of the signal, while it is strongly affected by the instrumental effects and the transmission through the atmosphere. The \textbf{fringe tracker} instrumental mode intends to ensure close loop operation and lock the fringe position on the detector (equivalent to adaptive optics module for single dish operation). The \textbf{coherencing mode} dedicated to open loop operation returns the position of the whole fringe packet when the guiding star is too faint to achieve a sufficient Signal to Noise Ratio (SNR hereafter) to operate in close loop. In both modes the smallest number of pixels is required in order to optimize fastest operation on faintest sources. An accurate sampling of the fringe pattern must be done, using a sufficiently long coherence length in order to catch a shifted fringe packet. The required coherence length allows the instrument to deal with disturbing effects that induce phase jumps and fringe smearing, and with contrast losses due to longitudinal chromatism.\\
A fringe tracker is operated with short exposures able to drive the servo-loop at the required speed (mainly given by atmospheric characteristic time $to$, down to  few millisecond). Coherencing mode may be operated at longer exposures time. The ultimate performances in both cases are achieved thanks to photo-counting detection that is able to deal with various temporal requirements of the incoming signal.

 \textbf{Spectral resolution} capability is mandatory for astrophysical investigations and to provide the appropriate sampling of the foreseen spectral lines. Up to $35 000$ resolutions are considered in current operated instruments \cite{mourard08, amb07}. A trade off  must be done between the spectral resolution and the limiting magnitude for a given observation.
 
 \textbf{Higher complexity imagery reconstruction} is ensured thanks to a larger number of baselines, for instance with a larger number of telescopes. In that case the beam combining instrument must deal with the corresponding number of beams, leading to rather complex instruments.Image reconstruction needs require also in this case to increase the number of combined telescopes and of coded baselines.
 
  \textbf{Low flux sensitivity} is required for the observation of the faintest objects. It can be achieved thanks to long exposure acquisition. In this case, an associated fringe tracker is mandatory to prevent from any fringe blurring during the exposure. However the faintest sources will be distinguishable using detectors limited by the photon noise. In that case individual long exposures are no longer required.
  
  \textbf{High photometric dynamic} intends to provide the detection and the analysis of faint structures in the vicinity of a bright sources (extra solar planets, surrounding disks). The detection of exo-earths requires at least a $10^6$ flux dynamic. The higher dynamic will be achieved using photon-counting detectors limited by their photo counting rate.
  
  Devoted designs must be chosen, according to the foreseen astrophysical science cases, but all of these specifications cannot be met with a single instrument.

\subsection{Integrated optics applied to interferometry}
Integrated optics (hereafter IO) developed for telecom applications is available in the $0.8-2.4 \mu m$ spectral range using various technologies. It provides very convenient toolbox is then available with suitable performances for its application to astronomical interferometry. A first review of IO capabilities applied to astronomical interferometry has been proposed ten years ago \cite{Kern96}. The most interesting capability of this technology is to integrate on a single tiny chip a few centimeters long an instrument with several optical functions made of complex optical circuitry. It yields very compact instruments, and brings high instrument reliability, with a low sensitivity to external constraints such as vibrations and temperature fluctuations. The intrinsic stability of an integrated instrument induces excellent phase behaviors. Integration and maintenance operations are highly simplified, while as soon as the fiber optics connection is properly managed the main part of the instrument is self aligned.

Many convenient optical functions can be implemented on the same chip, like light deflection, beam separation and coherent combination, modulation, light dispersion \cite{kern01,malbet99}. Furthermore, several coherent beam combination schemes are made available using single mode IO, including new schemes that are not provided by conventional bulk optics means. A description of the possible single mode IO beam combining schemes with their corresponding efficiency in terms of SNR have been proposed as a guideline for the designs of optimized components dedicated to on-going instrumental programs \cite{lebouquin04}. The three main schemes that have been considered and tested are multi-axial scheme (all in one combination) using spatial coding on the detector, co-axial one (pair wise combinations) using temporal coding thanks to a modulation of the optical path difference, and a so-called co-axial  ABCD concept (pair-wise combination) which provides a built-in phase coding on four outputs of the component.

An important behavior of IO components is its filtering capability while operating in single mode conditions. It allows a significant improvement of the interferometer precision, thanks to a reduction of the error on the contrast measurement \cite{lebouquin06,kervella04}. However it must be noticed that single mode operation limits the field of view of the interferometer to compact object observation. The field of view is limited to the diffraction pattern of the individual telescopes of the interferometric array, while the resolution of the reconstructed image is given by the length of its largest baseline.

The IO potential have been demontrated thanks to on-sky exploitation leading to astrophysical results using the IOTA interferometer (Mt Hopkins - Arizona ) \cite{kraus05, monnier04} and the VLTI. Further in progress developments consider the use of complex IO components as VSI (VLTI Spectro Imager) \cite{malbet08} and Gravity \cite{Eisenhauer08} for the second generation instrument of the VLTI.
For space borne applications, if the compactness is a trivial advantage, with a direct impact on the payload of the instrument, one may mention also the strong advantages of such integrated instrument. During the integration phases the constrains applied to the integration and alignment procedures is significantly reduced. Most of the efforts are reported during the design phase. It reduces the unreliability for the duration of these integration phases.

An important drawback of single mode IO so far, is the limitation of the field of view to the  diffraction Airy disk of the individual telescopes. It must be also pointed out that available IO technology limits its application to near IR single mode instrument (up to $2.5 \mu m$ ), even if some fruitful technological developments for the manufacturing of IO components working at longer wavelengths provide already fruitful results. Related R\&D developments are in progress driven by NASA and ESA for the preparation of the space mission dedicated to the observation of earth like planets \cite{Lawson08, Wallner08} that requires an operation in the $4 - 18 \mu m $ spectral range.

\subsection{Stationary-Wave Integrated Fourier Transform Spectrometer principle }

 \begin{figure}[!t]
   \begin{center}
   \begin{tabular}{c}
   \includegraphics[width=\textwidth]{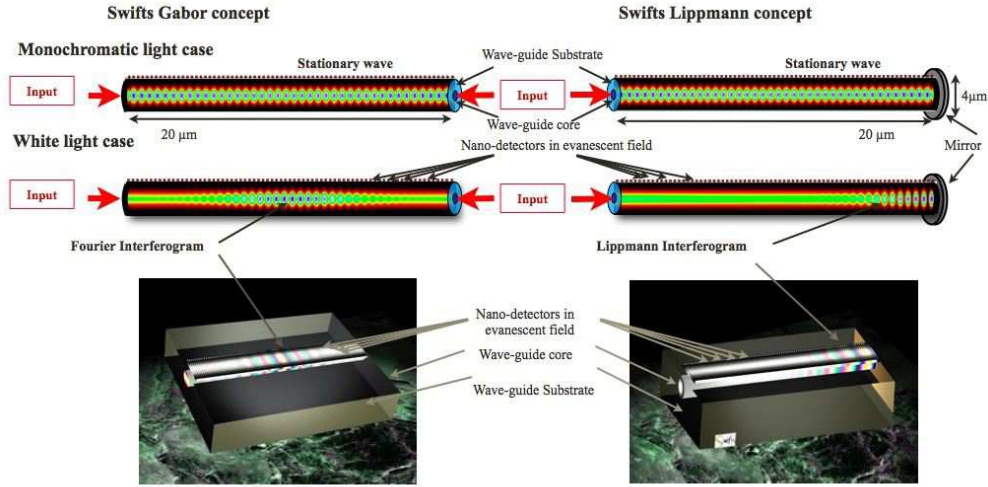}
   \end{tabular}
   \end{center}
   \caption[example] 
   { \label{fig:swifts-principle} 
\textbf{SWIFTS principle description:} Nanodetectors are placed in the evanescent field of the waveguide. Each detector samples a small part of the flux over a distance smaller than the fringe size. For polychromatic input light, the resulting superimposition of stationary waves led on to a Fourier interferogram sampled by the set of nanodetectors. The lower 3D artist views present the corresponding SWIFTS devices for colored entrance light and the resulting interferograms.\\
 \textbf{Left: SWIFTS-gabor principle.} The two coupled waves propagate in the waveguide in counter-propagative mode. The stationary waves are centered on the zero optical path difference location giving an access to the measurement of the interferogram phase.\\
 \textbf{Right: SWIFTS-Lippmann principle.} The forward propagating optical field coupled in the waveguide is reflected on the mirror at the waveguide end. The stationary waves are locked on the mirror leading to the so-called Lippmann interferogram starting by a black fringe .\\
}
  \end{figure}
    
The Stationary-Wave Integrated Fourier Transform Spectrometer (SWIFTS), encapsulate a new concept for on-chip spectro-detection (see Fig. \ref{fig:swifts-principle}): The detection is achieved along a standing waves within a single mode waveguide by probing its evanescent field, using nano-detectors. Based on this concept a motionless device is able to probe the spectral information of the incoming signal  \cite{LeCoarer07}. Several implementations have been investigated.

In a first mode named {\bf SWIFTS-Lippmann} (see Fig. \ref{fig:swifts-principle} right), the light is coupled into a wave-guide ended by a mirror. The standing waves form the interference pattern along the wave-guide, with its central black fringe locked onto the mirror. It led to a Lippmann interferogram that starts by a black fringe.  This mode gives only access to the spectra of the incoming light, like for a Fourier Transform Spectrometer.

In a second mode named {\bf SWIFTS-Gabor} (see Fig. \ref{fig:swifts-principle} left), the light is split in two equal parts thanks to a suitable IO divider or to a bulk optics beam splitter. These two parts are injected in counterpropagative scheme from the two extremities of a waveguide. In this scheme the two arms of the resulting interferometer are accurately balanced by construction in order to place the central fringe at the center of the detection zone. The resulting standing waves forms a Fourier interferogram along the guide. In this mode, in addition to the spectra of the incoming light, one has access also to its phase thanks to a measurement of the central fringe displacement with respect to the theoretical central fringe location. A design based on this mode is able to achieve interferometric metrology, thanks to a measurement of the fringe position.

 In both cases each element of a set of nano-scale detectors or intermerdiate diffusing dots placed in the evanescent field extracts only a small fraction of the guided energy. The pitch of the nano detector set matches a convenient sampling of the standing wave, preferably more than 2 samples per fringe. The nano-detectors must be smaller than $1/6$ of one fringe taking into account the refraction index of the propagating material. In order to maximize the signal the sampling is performed using a tiny square rather than a dot, that provides an average over its area. It turns into a contrast loss of 10\% for a  $1/4$ pitch.
  
 This approach based on a detection in the evanescent field, allows proper sampling of the interferogram using small size detectors in comparison to the quarter wavelength of the guided light.  It has been shown that an ideal positioning of the wires of the SWIFTS arrangement permits the spectrometer to build an interferogram using $74\%$ of the incoming photon \cite{LeCoarer07}. First prototypes provide a proper demonstration of the principle at various wavelengths. The above mentioned publication presents a measured spectrum over an 80 nm range ($1.5\mu m - 1.58\mu m$) with a 4 nm resolution.
 
\section{Proposed concepts}
\label{sec:swifts}
Having demonstrated the proper operation of IO coherent beam combiner \cite{berger01} and of spectro-detection in the vicinity of wave guides (SWIFTS) \cite{LeCoarer07}, we have the required building blocks for a new beam combiner design with on-chip detection of the fringes.
We propose here three possible schemes using demonstrated technologies.

\subsection{Toward Photo-counting Fringe trackers}
\label{sec:swiftsFT}
   \begin{figure}[!t]
   \begin{center}
   \begin{tabular}{c}
  \includegraphics[width=0.5\textwidth]{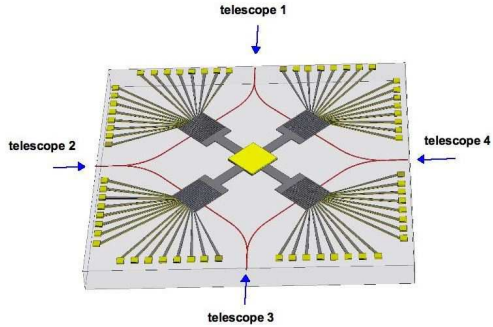}
   \end{tabular}
   \end{center}
   \caption[example] 
   { \label{fig:swifts-sspd} 
Beam combiner design for a 4 telescope configuration associated with SSPD detectors. Pair wise combinations are performed in contra-propagative mode in the waveguides located along the sides of the square (red lines on the figure). The set of 70 nm large SSPD wires allows to pick up the evanescent signal of the standing waves of this SWIFTS-Gabor configuration. In the proposed concept the 16 wires led to a resolution of 8, well adapted to fringe tracker applications.}
\end{figure}
  In this concept  (see Fig.  \ref{fig:swifts-sspd}) the interference fringes are located within the equalized waveguides that links the baseline of the interferometer.
We propose to implement a set of SSPD (Superconducting Single Photon Detector) \cite{feautrier07,verevkin02} wires in the vicinity of the waveguides which will sample and extract the convenient part of the evanescent field.

\begin{figure}[!h]
   \begin{center}
   \begin{tabular}{c}
  \includegraphics[width=0.9\textwidth]{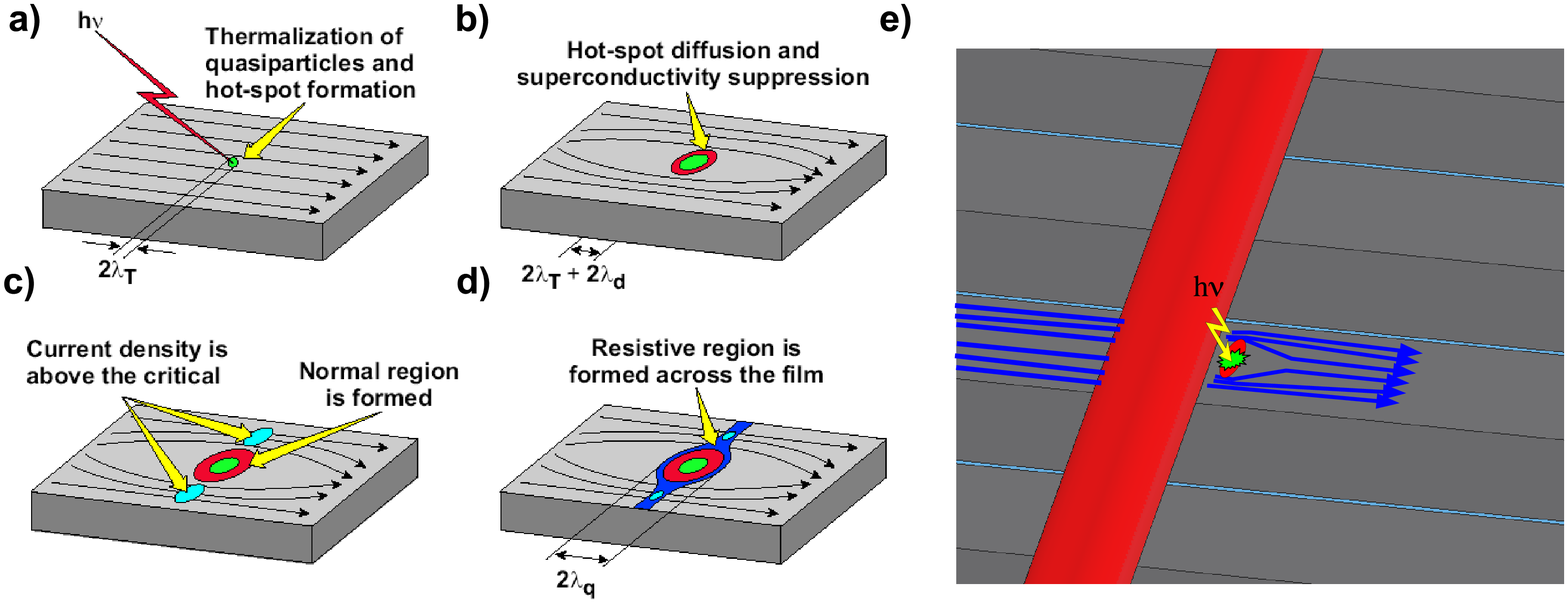}
   \end{tabular}
   \end{center}
   \caption[example] 
   { \label{fig:swifts-princ-sspd} 
Principe of the SSPD concept according to \cite{verevkin02}. Four states of a SSPD when a photon 
hits the superconducting materials where a subcritic current is installed: \\
 a) The photon is absorbed and heat locally the material.   b) The material is 
locally over the critical temperature inducing a larger resistivity. The current 
lines are deviated and the density of current c) becomes over the 
superconducting limit  d) Current is temporaly stopped and the heat is 
dissipated in the substrate and the line becomes again superconductive. \\
e) All SSPD wires are placed in the evanescent field of the waveguides (red strip of the figure). The photons
can disrupt any wires according to the intensity of the interferogram at 
crossing  location.}
   \end{figure}

The principle of SSPD is reported in Fig. \ref{fig:swifts-princ-sspd}. A DC current polarizes a wire of superconductor material, for example Niobium Nitride (NbN), just below the superconducting critical current. The interaction with a photon induces an elevation of the local temperature and a not null resistance of the wire close to the absorption point. The current lines are diverted and the local density of current flow reaches the critical value: the wires loose their superconductivity behavior inducing a voltage peak on the dipole. Due to a thermal effect, the NbN line returns to the superconducting state in a very short time ($ < 1~ns $) that allows to count high numbers of photons. Therefore, this detector presents demonstrated photon counting capabilities at a 1Gph/s rate with quantum efficiency as high as $50\%$ for the visible-near IR domain. Dating of incoming events can be performed with a 20ps accuracy. One driving application of SSPD is quantum cryptography.

The NbN wires are reported on a substrate surface like sapphire. While the implementation of optical waveguides on Sapphire substrate has been demonstrated\cite{Laversenne04}, it is possible to consider to report a suitable SSPD setup  on an integrated optics circuit designed for a multi telescope beam combination.

Figure  \ref{fig:swifts-sspd} shows a possible 4 telescope implementation. Each incoming beam is split to interfere with its the two nearest one within the optical waveguide. In the proposed scheme the standing waves are sampled using 16 nano wires leading to a resolution of 8. With 4 wires per fringe and considering an operation at $1.5 \mu m$ in Sapphire ( $n = 1.7$), the 16 wires $125 nm$  apart lie on a $2 \mu m$ length over a set of 4 fringes. The design is optimized in order to have the fringes centered on the wires array when all the optical path differences are properly compensated. This concept is able to provide the sensor of a close loop system to drive the suitable corrective delay lines.
An other design has been proposed which codes all the baselines in the 4 telescope configuration using diagonal guides in the square layout. Designs for larger number of telescopes can be proposed using the corresponding polygon shape in a circle scheme that code the minimum number of baselines required for a fringe tracker in a redundant mode.
The proposed concept requires specific development to obtain satisfactory behaviors of the combining waveguides at the temperature of operation of SSPD. NbN SSPD are operated at liquid Helium temperature sufficiently far from the superconductive critical temperature of the NbN film (11K), reducing the sensitivity to the operating temperature. Then SSPD are not very sensitive to temperature changes contrary to some others super conducting detectors as Transition Edge Superconducting detectors (TES) or Super conducting Tunnel Junction (STJ).
An extension of this design can be considered using other nanodetectors like QDIP (Quantum Dot Infrared Detector) for example or with new high temperature superconducting materials such YBaCuO which allow 77~K operation.

Suitable modeling using Rigorous Coupled-Wave Analysis (RCWA) \cite{morand08} shows that a global detecting efficiency of $30\%$ is achievable in such a case. In these conditions considering the  beam combination of $2\times 8 m$ telescope ( VLTi ) and a SNR of 3, we calculate that fringe tracking can be performed at 1kHz rate for a stellar magnitude K=14 (Stellar magnitude for the astronomical photometric K band ($2 \mu m - 2.4\mu m$). The photo counting mode brings even faster operation. It brings also the capability to deal with a very high dynamic of the incoming flux, mainly limited by the maximum speed of the detector electronics.

This technology is well suited for fringe tracker applications where only a limited number of sampling point is required to sense the phase of the interferometric signal. It led to a SWIFTS-Gabor concept with only a limited number of nano-wires. It potentially provides photo-counting capability at very high speed. The photon counting process at very high speed ensures also a sampling of the incoming flux over an extended photometric dynamic.

Current SSPD developments in progress in a LAOG-CEA collaboration aims at preparing elementary built-in blocks of such a concept.

\subsection{Toward an integrated  instrument with high spectral resolution capability}
   \begin{figure}[!t]
   \begin{center}
   \begin{tabular}{c}
   \includegraphics[height=10cm]{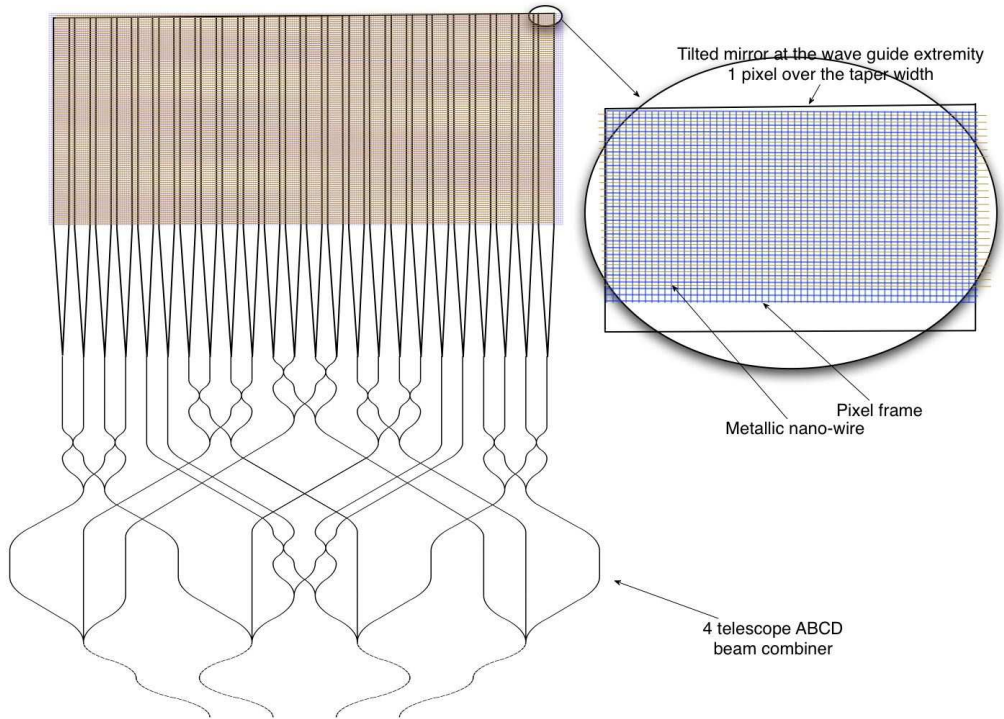}
   \end{tabular}
   \end{center}
   \caption[example] 
   { \label{fig:swifts-abcd} 
A 4 telescope beam combiner implementation using conventional FPA. Each output of this pair wise beam combiner is sampled within an enlarged guiding area. All enlarged guide ends with a tilted mirror surface that induces a tilted SWIFTS-Lippmann interferogram. A set of parallel nano-metric wires pick up a part of the signal in the evanescent field of the planar guide.}
   \end{figure} 

For this setup (see Fig.  \ref{fig:swifts-abcd}) we propose a modified design of an existing chip dedicated to the combination of the beams coming from 4 telescopes in matrix mode in the near infrared \cite{Jocou08,lacour08}. In this design the optical layout provides on its outputs a set of 4 phases status without any temporal modulation , which allows the proper sampling of the interference pattern. The sampling of all available baselines lead to a 24 output chip.

For all output guides, we propose to implement an enlarged waveguide zone ended by a slightly tilted integrated mirror. This mirror induces tilted standing waves in the enlarged zone where a set of nano-wires picks up the evanescent field above the wave-guides. An IR Focal Plane Array (FPA hereafter) is mounted beneath this enlarged zone performing the detection of the SWIFTS-Lippmann fringes able to measure the output spectrum at each channel output.  The FPA pixel lines are located in coincidence with the wire arrangement.  The resulting standing waves in the near IR ($ \lambda = 1.5\mu m , and ~ n=1.5$) present a wavelength of 500nm, leading to an undersampling using $18.5\mu m $ pixels. The proper sampling will be performed as soon as the enlarged guide width is sampled by 74 pixels, as each pixel line is surrounding 37 fringes. 

This configuration is well-suited for high spectral resolutions since this one is given by the length of the enlarged zone surrounded by the FPA. For example an Hawaii2 FPA with a line of $1024 \times 18.5 \mu m$ pixel provides a spectral resolution of R=40~000 at $1.5 \mu m$ (e.g. 0.04 nm).
Among the proposed concepts in this paper, it is certainly the cleanest one while it provides an accurate measurement over all the baselines with an identical number of samples.
   \begin{figure}[!h]
     \centering \includegraphics[width=0.9\textwidth]{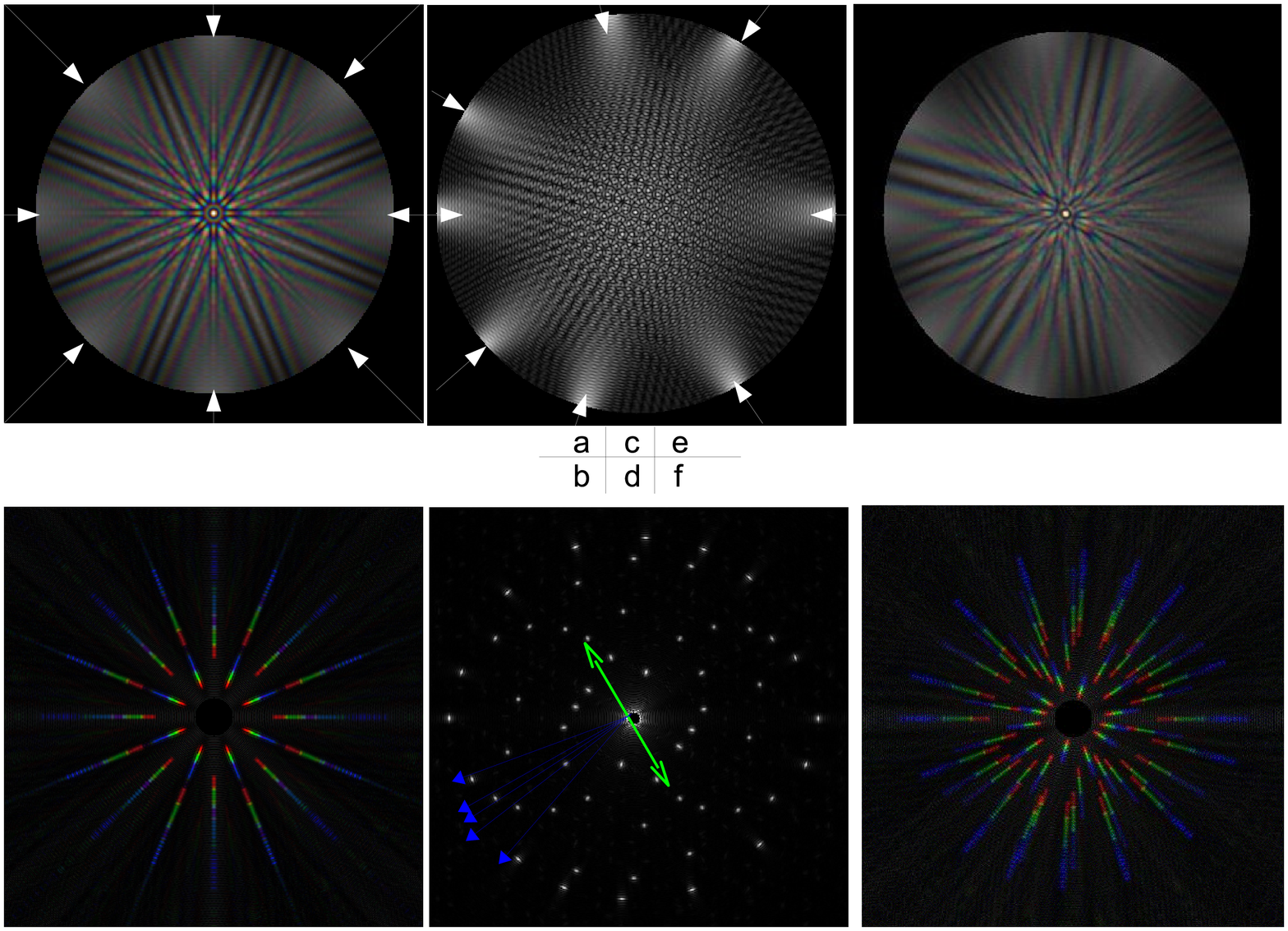}
   \caption[example] 
   { \label{fig:swifts-extd-model} 
Simulation for a Fizeau-Narcisse configuration for an 8 T beam combiner.\\
The upper part shows the SWIFTS-Gabor pattern (holographic pattern) in the planar waveguide fed by 8 incoming waveguides.\\ 
The lower part corresponds to its Fourier transform, giving the power spectra along all available baselines.\\
The middle part corresponds to the monochromatic case (c, d), while the right and left part corresponds to three chromatic bands (150nm wide) covering the visible domain that illustrates the chromatic behavior of the component(a, b, e, f). \\
One baseline (2 inputs) built up a fringe pattern that corresponds in the Fourier plane to a set of two fringe peaks in the monochromatic case (shown by the green arrow of the figure) or two symmetrical  spectra in the polychromatic case. The blue arrows show the individual spectrum directions.\\
\textbf{ a) and b)}: The waveguide inputs  pointed by the white arrows are regularly disposed around the circular beam combining planar waveguide. The corresponding Fourier transform illustrates how the spectra overlap inducing an information confusion (see Fig. \ref{fig:swifts-extd})\\
\textbf{c) d) e) and f)}: The waveguide inputs are managed in a non-redondant arrangement to avoid any overlapping of the resulting pairwise spectra.
}
   \end{figure} 
Such design can be implemented using modified available IO components associated with commercial FPA, bounded with the suitable nano-wire comb thanks to available technology platform.

Using the assumptions of section \ref{sec:swiftsFT}, one can derive the achievable sensitivity of an instrument based on this concept. It has been demonstrated \cite{LeCoarer07} that a set of nano wires properly designed is able to extract up to $74\%$ of the incoming signal for an optimized setup. A FPA mounted in front of these wires is able to detect half part of the diffused light, leading to a $37\%$ efficiency. In a near future one can consider to use a close to noiseless IR detector \cite{Rothman08}, bounded on the above described near IR IO beam combiner. Under conservative assumptions one may consider that the foreseen IO combining component present an efficiency of $30\%$. Considering a resolution of 40~000 over 4 outputs with an exposure time up to $120 s$, using a SNR of 10, such an instrument will be able to attain a theoretical stellar magnitude $K=14.3$.

A similar development for spectroscopy purposes in an other spectral range, using an arrangement of parallel straight waveguides is currently under progress by the authors funded by a FUI program of the French government (see Acknowledgments).  The adaptation for the applications to interferometry requires the use of a suitable IO beam combiner.

\subsection{Extended Planar concept }

   \begin{figure}[!t]
     \centering \includegraphics[width=0.6\textwidth]{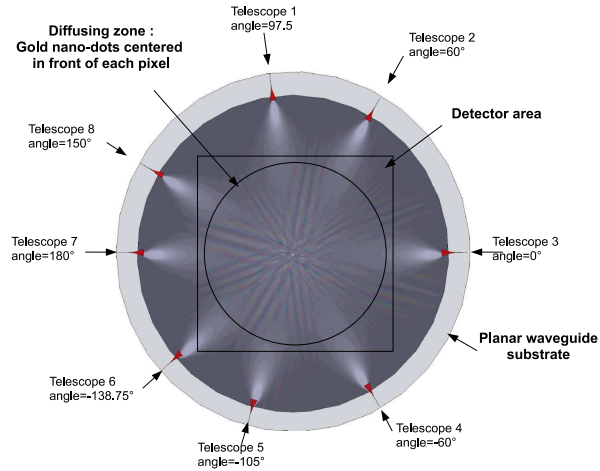}
   \caption[example] 
   { \label{fig:swifts-extd} 
Implementation of the Extended Planar Concept. A proper arrangement of gold nano-dot is implemented on the planar wave guide fed by the tapered input waveguides. This IO component is directly mounted on a FPA, with a pixel corresponding to each diffusing dot.}
   \end{figure} 

The concept (see Fig.  \ref{fig:swifts-extd}) is based on an extension of the SWIFTS-Gabor principle where the standing waves resulting from a large number of inputs are managed in a planar waveguide. It provides sophisticated signal coding, using a simple implementation. The data processing requires a complex appropriate Fourier treatment. A planar waveguide is fed by all incoming beams (8 in the case of  Fig. \ref{fig:swifts-extd-model}), brought thanks to single mode waveguides all around the beam combining zone. The arrangement of the inputs all around the circle, must not introduce common bisections to avoid any signal overlapping in the resulting Fourier inversion ( see Fig.  \ref{fig:swifts-extd-model}) . For instance a regular disposition of the input guides would produce a fringe pattern with equivalent coding frequencies for all baselines, and their contributions would not be distinguishable.  A tapered zone is implemented at the interface of each incoming wave-guide, in order to manage the diffraction effects within the planar wave-guide and to generate a pure plane wavefront and to ensure proper use of Fourier inversion. For the case of spherical wavefront, the inversion law must be obtained using an interaction matrix calibration. It must be noticed that this setup can be operated with all or part of the 8 inputs. For instance, if some of the telescopes are not available, the instrument can be still operated with induced losses of performances related to the missing flux from the corresponding telescopes.

A matrix of nano-dots spread out the light from the resulting SWIFTS-Gabor fringes, collected by the pixels of a near IR FPA bounded at the waveguide surface (see Fig. \ref{fig:swifts-extd}).

The pixel pitch of available FPA (around 20 $\mu m $ for near IR HgCdTe detectors) induces an under-sampling condition while the fringe  width inside the waveguide material is almost 500~nm at $1.5\mu m $. To ensure proper complete sampling of the SWIFTS-Gabor pattern it is considered either to modulate the Optical Path Difference for all baselines, or to take benefit of the high redundancy available in the resulting interferometric pattern. The efficiency performances of this setup will be given by the performances of the associated FPA.

This last concept illustrates the level of integration that can be achieved using SWIFTS principle. One difficulty to implement this concept resides in the bounding process between a planar waveguide and a regular FPA through a nano-dot matrix. This concern is currently addressed through the above mentioned  R\&D program. The performance evaluation of a setup based on this concept requires further analysis, but it can be expected that it should be more efficient than the previous one, since it requires simpler optical circuitry and take benefit from the same detection process.

Another concern is the processing of the interferometric signal to reconstruct the observed images. Related signal processing developments to address this issue are starting at LAOG. Such a setup is able to handle a large number of inputs, providing that the number of pixels is sufficient to ensure the sampling of the interferometric pattern at the required resolution. Large spectral resolution is intrinsic to the concept.

\label{sec:title}

\section{Conclusion}
 
In this paper we propose some possible implementation of SWIFTS-based coherent beam combiners, for single mode interferometry, using technologies   which are available or under development. Such promising integrated instruments may provide both compact setups with all deriving behaviors. Furthermore the concept provides optimized performances as it is possible to consider adapted designs according to the foreseen instrumental and astrophysical application. The design allows to meet most of the design requirements for an instrument dedicated to stellar interferometry.

The single mode operation and the compactness of an integrated instrument ensures the measurement precision. On-chip detection ensures accurate and intrinsically stable sampling of the fringe patterns and provides an efficient phase measurement. Some possible concepts are able to manage a large number of sampling points opening the way to high spectral resolution and to large number of telescopes beam combination. The possibility of implementing a photo counting detection allows to consider high photometric sensitivity and high photometric dynamic together with high speed operation.

The introduction of SWIFTS concept, opens the way  for the design of
fully integrated interferometric instruments that include sophisticated 
functions such as phase modulation, junctions, switches , spectrometry and 
detection. Even if other photonics spectrometer concepts can also be envisaged to improve 
the signal to noise of such instruments, due to its intrinsic simplicity only 
SWIFTS based concept are currently under development for an application to astrophysical interferometry.


 \section*{ACKNOWLEDGMENTS}       

This work has been supported thanks to CNES funding. A support was also alocated to the project by the ASHRA / INSU from CNRS.\\
On going SWIFTS developments are led by Thierry Gonthiez (Floralis / http://floralis.fr/us/), mainly for the program labeled by the global competitiveness cluster MINALOGIC and funded through the national french FUI program (Fonds Unique Interminist\'eriel).\\
The authors would like to thanks Jean Philippe Berger and Fabien Malbet for their fruitful remarks on this manuscript.
SSPD developments are conducted by Philippe Feautrier at LAOG together with Jean Claude Vill\'egier at CEA-Grenoble.



\end{document}